\documentstyle[12pt,aasms4]{article} 
\def\etal{{et~al.}\ }

\def\kpc{{\rm\,kpc}}

\def\kms{{\rm\,km/s}}

\def\msun{{\rm\,M_\odot}}

\def\vol#1  {{{#1}{\rm,}\ }}

\def\lya{{\rm Ly}\alpha}
\def\mytau{\tau_{{\rm Ly}\alpha}}

\def\etal{et al.\ }

\def\clock{\count0=\time \divide\count0 by 60
     \count1=\count0 \multiply\count1 by -60 \advance\count1 by \time
     \number\count0:\ifnum\count1<10{0\number\count1}\else\number\count1\fi}

\begin{document}
\title{On the Clustering of Lyman Alpha Clouds}
\author{Renyue Cen\altaffilmark{1}, Steven Phelps\altaffilmark{2},
Jordi Miralda-Escud\'e\altaffilmark{3} and
Jeremiah P. Ostriker\altaffilmark{4}}

\altaffiltext{1} {Princeton University Observatory, Princeton University, Princeton, NJ 08544; cen@astro.princeton.edu}
\altaffiltext{2} {Department of Physics, Princeton University, Princeton, NJ 08544; phelps@astro.princeton.edu}
\altaffiltext{3} {Department of Physics and Astronomy, University of Pennsylvania; jordi@llull.physics.upenn.edu}
\altaffiltext{4} {Princeton University Observatory, Princeton University, Princeton, NJ 08544; jpo@astro.princeton.edu}

\begin{abstract}
We examine the correlational property of $\lya$ 
clouds (along the line of sight)
in detail utilizing a hydrodynamic simulation of 
$\lya$ clouds in a cold dark matter universe with
a cosmological constant,
and compare it to that of mass and galaxies.
We show that the correlation strength of $\lya$
clouds is somewhat weaker than that of the underlying matter,
which in turn should be weaker than that of galaxies (biased
galaxy formation).
On the scales probed, $10-300$km/s, for $\lya$  clouds
we find that
higher density, higher optical depth,
higher column density regions are more strongly
clustered than lower density, lower optical depth, lower column density
regions,
with the difference being larger at small separations and
smaller at large separations.
Thus, a consistent picture seems to emerge:
the correlation strength for a given set of objects is positively 
correlated with their characteristic global density
and the differences among the correlations of galaxies, $\lya$ clouds
and mass reflect the differences in density that each trace.
Significant positive correlations with a 
strength of $0.1-1.0$ are found for $\lya$ clouds
in the velocity range $50-300$km/s.
This effect should be observable.
The correlation function of $\lya$ clouds seems
to be a monotonically decreasing function of separation,
indicating that correlation strength
should be less than $0.1$ at $\Delta v>300$km/s,
where our current simulation box is too small to give a
reliable measure.

Among the correlational measures examined, 
an optical depth correlation function
(Equation 5) proposed here
may serve as the best correlational measure.
It reasonably faithfully represents the true correlation of the underlying
matter, enabling a better indication
of both matter correlation and the relationship between galaxies
and $\lya$ clouds.
Furthermore, it appears to be an 
alternative to the conventional line-line correlation function
with the virtue that
it does not require ambiguous post-observation
fitting procedures such as those commonly employed 
in the conventional line-finding methods.
Neither does it depend sensitively on the observational resolution
(e.g., FWHM), insofaras the clouds are resolved
(i.e., the FWHM is smaller than the line width).
Conveniently,
it can be easily measured with the current observational
sensitivity without being contaminated significantly by the
presence of noise, if one chooses an appropriate optical
depth floor value $\tau_{min}$ (an adjustable parameter)
say, $\le 2.0$.

\end{abstract}


\section{Introduction}

With rapid advances in observational capabilities in recent
years, thanks to the advent
of the ground-based 10m Keck telescopes  as well
as the Hubble Space Telescope,
observation of the 
$\lya$ forest has become
a highly useful tool for probing
the universe at low-to-moderate redshift
(e.g., Carswell \etal 1991; Rauch \etal 1992; Petitjean \etal 1993;
Schneider \etal 1993; Cristiani \etal 1995; Hu \etal 1995;
Tytler \etal 1995; 
Lanzetta \etal 1995; Bahcall \etal 1996).
When analyzed in conjunction with progressively more sophisticated
and accurate theoretical modelling 
(Cen \etal 1994, CMOR hereafter; Zhang \etal 1995;
	Hernquist \etal 1996;
Miralda-Escud\'e \etal 1996, MCOR hereafter),
made possible by the availability of powerful computers and
accurate cosmological hydrodynamic codes,
observations of $\lya$ clouds should 
greatly enhance our ability to test cosmological
models.
The above quoted papers show such a good agreement between theory
and observation with regard to physical properties
which have already been documented,
that we are led to 
use the theory now to predict properties,
such as spatial correlations among 
$\lya$ clouds, that have not yet been reliably measured.

Early observations that $\lya$ clouds are very weakly clustered
in velocity space on scales $\ge 300$km/s along the line of sight
to distant quasars
at high redshift (Sargent \etal 1980),
compared to the expected correlations among galaxies at the
same redshift,
provided the first compelling evidence that $\lya$ clouds
at high redshift are perhaps not associated with galaxies.
This single piece of evidence, combined with
the inferred temperature of the clouds,
has spawned most subsequent
discussions about their origin and nature 
(Ostriker \& Ikeuchi 1983; Ikeuchi \& Ostriker 1986; 
Rees 1986; Ikeuchi 1986; Bond, Szalay, \& Silk 1988).
On the other hand, the recent observations by HST
seem to indicate that $\lya$ clouds at low redshift
($z\le 1.0$) may have their origin directly inside
or in very close proximity to galaxies 
(Lanzetta \etal 1995; Bahcall \etal 1996).
Of course, ground based moderate redshift observations
and space based low redshift observations may sample
essentially different populations,
since the high redshift population is declining with 
decreasing redshift at such a large rate as to make
a negligible contribution to
the HST observed lines.
Moreover, 
various studies of the clustering of
$\lya$ clouds at high redshift, which primarily measure
line correlations along the line of sight,
have arrived at different results.
For example, Webb (1986) suggests that $\lya$ clouds are
weakly clustered with $<\xi>=0.32\pm 0.08$
over the velocity range $50-290$km/s,
while Rauch \etal (1992)
find no clustering at any scale
for the clouds as a whole.
Such seemingly conflicting observational evidence
at high redshift and between high and low redshifts
indicate that our current understanding 
of the clustering properties of $\lya$ clouds may be quite
incomplete.

In this paper we attempt to understand better the spatial
clustering 
properties of $\lya$ clouds by analyzing
the simulated $\lya$ clouds in the cold dark matter
model with a cosmological constant ($\Lambda$CDM).
This model, according to the MCOR analysis,
provides a good fit to the observed distributions of absorption
in the $\lya$ forest as measured by the one-point
distribution function of transmitted flux, namely the flux 
distribution and the derived column density distribution, 
but that work did not examine 
the two-point correlation of the $\lya$ clouds in much detail.
In the present work
several clustering measures are used,
including a line-line two-point correlation function and measures
(some newly proposed) directly using flux and optical depth distributions 
as well as more fundamental matter correlations, 
directly computable in simulations.
The paper is organized as follows:
A brief description of the simulation is given in \S 2.
The correlational measures used to analyze the simulation
are defined in \S 3.
Results and conclusions are given in \S 4 and \S 5.

\section{Simulation}

We simulate the formation of $\lya$ clouds
in a spatially flat cold dark matter universe
with a cosmological constant ($\Lambda$CDM),
with the following cosmological parameters:
Hubble constant $H_o=65$km/s/Mpc, 
$\Omega_{0,CDM}=0.3645$, $\Lambda_0=0.6$,
$\Omega_{0,b}=0.0355$ (cf. Walker \etal 1991), 
$\sigma_8=0.79$
(the simulations we use in this paper
are the same as those used in CMOR and MCOR).
The primary motivation for choosing this model
is simple: overall it provides an excellent fit to available
observations (Ostriker \& Steinhardt 1995, "Concordance model").
The simulation box size is $10h^{-1}$Mpc comoving
with $N=288^3$ cells and $144^3$ dark matter particles.
The cell size is $35~h^{-1}$kpc (comoving) corresponding
to an average baryonic cell mass of $6.3\times 10^5\msun$,
with the true spatial and mass resolutions being 
about 2 and 8 times worse than those values, respectively.
At $z=3$,
the Jeans length, $\lambda_J\equiv (\pi c_s^2/G\bar\rho_{tot})^{1/2}$
for $c_s=v_{rms}=10\kms$, 
is equal to $400 h^{-1} \kpc$ in comoving units, or 11 cells.
The power spectrum transfer function is computed
using the method described in Cen, Gnedin, \& Ostriker (1993).
We use a new shock-capturing 
Total Variation Diminishing (TVD) cosmological hydrodynamic code 
described in Ryu \etal (1993).

All the atomic processes for a plasma of (H, He) of primeval
composition (76\%,24\%) in mass are 
included, using the heating,
cooling, and ionization terms described in Cen (1992).
We calculate self-consistently
the average background photoionizing
radiation field as a function of frequency, assuming the radiation field
is spatially uniform (i.e., optically thin).
 The evolution
of the radiation field is calculated given the average attenuation in the
simulated box and the emission (both from the gas itself and from the
assumed sources of ionizing photons).
The time-dependent equations for the ionization structure of the gas
are solved by iteration using an implicit method, to avoid the
instabilities that arise in solving stiff equations. 
In general, the
abundances of different species are close to ionization equilibrium 
between recombination and photoionization
after most of the gas has been photoionized.

We model galaxy formation as in Cen \& Ostriker (1992, 1993a,b).
The material turning into collisionless particles as 
``galaxies" is assumed to emit ionizing radiation 
with two types of spectra:
one characteristic of star formation regions and the other
characteristic of quasars, with efficiencies (i.e., the fraction of
rest-mass energy converted into radiation) of
$e_{UV,*}=5\times 10^{-6}$, and $e_{UV,Q}=6\times 10^{-6}$, respectively.
We adopt the emission spectrum of massive stars from Scalo (1986) 
and that of quasars from Edelson and Malkan (1986).
These coefficients have been chosen to produce
a computed radiation field consistent with that obtained from
the proximity effect and other observational constraints.
Details of how we identify galaxy formation and
follow the motions of formed galaxies
have been described in Cen \& Ostriker (1993a).
Note that in this simulation
supernova energy feedback
from aging massive stars 
into the intergalactic medium is not included.

\section{Correlation Measures}

We analyze the clustering properties of $\lya$ clouds
along the line of sight at $z=3$
using six different two-point correlation functions:
$\xi_{\rho_d}(r)$, 
$\xi_{\rho_b}(r)$, 
$\xi_{flux}(v)$, 
$\xi_{\tau,c}(v)$, 
$\xi_{\tau,s}(v)$, 
and $\xi_{line}(v)$. 
They are defined as follows.

\begin{equation}
\xi_{\rho_x}(r)\equiv {<\rho_x(r+r^\prime)\rho_x(r^\prime)>\over <\rho_x>^2}-1\quad ,
\end{equation}
\noindent where $\xi_{\rho_x}(r)$ is the two-point
correlation function (Peebles 1980)
of the cold dark matter density ($x=d$)
or baryonic density ($x=b$);
$\rho_x(r)$ is the density at point $r$ along the line
of sight; 
$<\rho_x>$ is its mean averaged over 
all lines of sight sampled.
The angular brackets in 
the numerator indicate averaging over all
pairs of pixels with the line of sight separation between
$r$ and $r+\Delta r$.
Note that for $\xi_{\rho_d}(r)$ and  
$\xi_{\rho_b}(r)$ 
we ignore the peculiar velocity along the line of sight
so that the resulting correlations reflect
the more fundamental matter correlations in real space.
This fundamental correlational measure will be
compared with other measures, which are directly observable
and calculated in velocity
space, including the peculiar velocity as well as the thermal effect.

The two-point correlation function of the normalized transmitted flux $F(v)$
along the line of sight, 
$\xi_{flux}(v)$, is defined by
\begin{equation}
\xi_{flux}(v)\equiv {<F(v+v^\prime) F(v^\prime)>\over <F>^2}-1\quad ,
\end{equation}
\noindent where $<F>$ is its mean averaged over 
all lines of sight sampled and
the angular brackets in 
the numerator indicate averaging over all
pairs of pixels with the line of sight separation between
$v$ and $v+\Delta v$.
Note that this flux correlation measure
has a slightly different definition than
another flux correlation measure used in MCOR
(see equation 2 of MCOR for definition).

Next, we propose a new two-point
correlation function of (modified) optical depth
along the line of sight, defined as
\begin{equation}
\xi_{\tau,c}(v)\equiv {<\tau_c(v+v^\prime) \tau_c(v^\prime)>\over <\tau_c>^2}-1\quad ,
\end{equation}
\noindent 
$<\tau_c>$ is its mean $\tau_c$ averaged over 
all lines of sight sampled
the angular brackets in 
the numerator indicate averaging over all
pairs of pixels with the line of sight separation between
$v$ and $v+\Delta v$.
$\tau_c(v)$ is {\it not} the optical depth as observed,
but is the latter modified as:

\begin{equation}
\tau_c(v)\equiv \hbox{Min}[\tau_{max}, \tau_{obs}(v)]\quad ,
\end{equation}
\noindent where $\tau_{obs}(v)$ is what is directly measured observationally,
related to the normalized transmitted flux by
$\tau(v)=-log[F(v)]$;
Min$()$ is the minimum operator, 
which takes the smaller of its two arguments.
The letter "$c$" in the subscript of
$\tau_c$ and $\xi_{\tau,c}$ is intended to stand
for "ceiling", imposed on $\tau_{obs}$ (equation 4).
Because of this notation,
we may call 
$\xi_{\tau,c}$ ``capped optical depth correlation function"
We use three values of $\tau_{max}=(2.0,4.0,\infty)$.
Note that in the case $\tau_{max}=\infty$, $\tau_c(v)=\tau_{obs}(v)$.
The purpose of choosing several $\tau_{max}$'s
is to study the dependence of clustering strength on
regions with different optical depth, different column densities.

We propose another new two-point
correlation function of (yet another modified) optical depth
along the line of sight, defined as
\begin{equation}
\xi_{\tau,s}(v)\equiv {<\tau_f(v+v^\prime) \tau_f(v^\prime)>\over <\tau_f>^2}-1\quad ,
\end{equation}
\noindent where $\tau_f(v)$ is defined as:

\begin{equation}
{\tau_f(v)\equiv \cases{1,&if $\tau_{obs}(v)\ge \tau_{min}$\cr
0,&if $\tau_{obs}(v)<\tau_{min}$\cr}}
\end{equation}

\noindent We pick three values of $\tau_{min}=(1.0,2.0,4.0)$,
for a similar intention
to study the dependence of clustering strength on
the different regions 
but in a somewhat different angle
than $\xi_{\tau,c}$.
The letter "$s$" in the subscript of
$\tau_f$ and $\xi_{\tau,s}$ is intended to stand
for "step", imposed on $\tau_{obs}$ (equation 6).
We will call 
$\xi_{\tau,s}$ ``stepped optical depth correlation function".

Finally, we test the usual line-line 
two-point correlation function along the line of sight, defined as
\begin{equation}
\xi_{line}(v)\equiv {\hbox{\# of pairs with } v \rightarrow v+\Delta v \over
\hbox{\# of pairs for a random sample with } v \rightarrow v+\Delta v}-1 \quad ,
\end{equation}
We use the method described in MCOR 
to identify lines.
This method is different from what 
is used in extracting lines from 
observed spectra, so a direct comparison 
between the simulation and observation is not possible
at this time.

In all cases the global average $<>$ is understood 
to be the average over our computed 
sample at a given redshift $<>_z$.
The purpose of examining several measures of correlation 
in the $\lya$ absorption spectra
is threefold.
First, we wish to understand the relationship between
correlations that are directly measured from observed absorption spectra
and the fundamental correlation of the underlying matter distribution.
For this purpose, we have included 
$\xi_{\rho_d}(r)$
and $\xi_{\rho_b}(r)$.
Second, we attempt to find 
a correlational measure which is relatively 
free from post-observation fitting procedures such as
line profile fitting and line deblending, and is relatively
insensitive to small changes in observational accuracy in terms
of signal to noise ratio and FWHM.
For this reason, we have calculated
$\xi_{flux}(v)$, 
$\xi_{\tau,c}(v)$ and  $\xi_{\tau,s}(v)$. 
Each of these three measures is directly based on the
observed flux distributions and does not require 
any fitting procedure.
Furthermore, we expect that, once the resolution 
of observations becomes higher than the width
caused by the expected thermal broadening,
these measures will be rather robust.
This property is in sharp contrast with that of line correlations,
for which ambiguity exists in deblending lines 
regardless of how high a signal to noise ratio an observation
can achieve.
Third, we want to understand the implications
of available observational results
obtained by employing the line-line two-point
correlation function. 
Therefore, we also examine the simulation using line-line
two-point correlation function $\xi_{line}(v)$, 
although a direct comparison with observations is still not possible
due to differences between the line identification algorithms used
in observations and in simulations.

For the first two correlations, 
$\xi_{\rho_d}$ and
$\xi_{\rho_b}$,
no observational noise is added
and their validity in a specific cosmological model
only depends on the accuracy of the simulation, 
which is primarily limited by
the simulation resolution and box size
(and of course additionally by the residual uncertainty
in the underlying cosmological model).
For the remaining four correlational measures
we generate two sets of spectra
with (S/N=10, FWHM=30km/s) and 
(S/N=50, FWHM=6km/s).
A sampling bin of $2$km~s$^{-1}$ is used in all cases.
Since we are interested in regions with large
amount of absorption ("clouds") as well as those with mild
amount of absorption (weak lines or "voids"),
the widely used Gaussian statistic of noise is not valid,
in general. 
We use the Poisson statistic of noise, which is valid
in both low and high flux regions assuming that
we know the signal to noise level at each pixel of 
a spectrum and is identical 
to the Gaussian statistic in the limit of large photon numbers.
This is done in the following manner.

First, we require that
the average flux decrement in the simulated spectra,
$\langle D\rangle=1-\langle \exp^{-\mytau}\rangle$,
be equal to the observed value, $\langle D\rangle_{obs}=0.36$ 
at $z=3$ (\cite{prs93}).
This is achieved by adjusting the parameter,
$\mu\equiv {(\Omega_{0,b}/0.015h^{-2})\over (h~j_{H}/10^{-12} \hbox{sec}^{-1})^{1/2}}$ (MCOR),
where $h$ is Hubble constant in units of $100$km/s/Mpc,
$\Omega_{0,b}$ is the mean baryonic density and
$j_{H}$ is the hydrogen photoionization rate.
This normalization procedure for the
flux distribution is {\it necessary} in order to fix its overall amplitude,
due to the large uncertainties of the observed values of
$\Omega_{0,b}$ and $j_H$.
The fitted value of $\mu$ is $1.90$ for the $\Lambda$CDM model.
Second, we add noise to the normalized spectrum.
By definition, the signal to noise ratio at the continuum is 

\begin{equation}
S/N={N_{src}\over \sqrt{N_{src}+N_{noise}}}.
\end{equation}

\noindent
where $N_{src}$ and $N_{noise}$ are 
the number of source photons at the continuum and
the number of noise photons,
respectively, per frequency bin.
Thus, given S/N and $N_{noise}$,
we can solve equation (8) to get $N_{src}$.
To simplify the illustration 
(without loss of generality) we assume that 
$N_{noise}$ is dominated by the detector read noise.
This is a good approximation only for bright quasars
where the number of sky photons is small
(due to a shorter exposure time) compared to
the read noise of, say, a CCD detector.
For example, the gain of the HIRES CCD detector
on the Keck telescope is 6.1 electrons, so the number of photons due to the
CCD read noise integrated over 5 spatial pixels (for each frequency bin)
is $N_{CCD}=5\times 6.1^2=186$.
For a $V=16.5$ mag quasar at $5000$A with 1-hr integration time,
there are about 4 photons from the sky per spatial pixel,
giving a total count of sky photons per frequency pixel (integrated
over the 5 spatial pixels) of only 20 photons
(see, e.g., Hu \etal 1995).
A frequency bin in the synthetic noise-free spectrum with flux $f$
contains $f N_{src}$ photons.
When noise is added,
the ``observed" number of photons (subtracted by the
known CCD read noise) in the frequency bin is

\begin{equation}
N_{obs} = Poisson(f N_{src} + N_{CCD}) - N_{CCD},
\end{equation}

\noindent where $Poisson(X)$ means a Poisson distributed random number
with the mean equal to $X$.
The $N_{CCD}$ value of the HIRES CCD 
detector is adopted in the calculations.

\section{Results}

For each correlation function computed we use 12,000 lines of sight
each having a length of $1280$km/s.
Figure 1 shows 
$\xi_{\rho_b} (r)$ (solid curve)
and
$\xi_{\rho_d} (r)$ (dashed curve).
Note that the two correlations 
are computed in real space as a function of proper distance,
as shown in the bottom horizontal axis.
For reference the velocity distance,
converted from the proper distance
using the relation $v=r~H(z)$,
where $H(z)=512$km/s/Mpc is the Hubble constant at $z=3$
for the adopted model, is shown in the top horizontal axis.
Note that $1+\xi$ rather than $\xi$ is shown in the plot
in order to display negative values of $\xi$ on a logarithmic scale.
We see that both dark matter and gas
are highly clustered at scales $\le 0.1h^{-1}$Mpc comoving,
with dark matter being more strongly clustered than
gas on all scales.
This apparently surprising result
that baryonic gas density is {\it anti}
biased with respect to dark matter
density has been found in all simulations to date begining
with our earliest numerical work (Cen 1992).
While no completely satisfactory explanation
has been given for this fact it presumably reflects
the greater instability of the dark matter to waves 
perpendicular to the compression vector.
The correlation lengths of the dark matter and gas,
defined as the length at which $\xi$ is unity,
are $0.63h^{-1}$Mpc and $0.46h^{-1}$Mpc comoving,
respectively.
The difference in the clustering between dark matter and gas
is enhanced due to the thermal photoheating effect in the baryons.
It is noted that the resolution of the simulation
is $0.087h^{-1}$Mpc (2.5 cells), 
so numerical smoothing in the gas can not be responsible for
the difference seen on scales $\ge 0.1h^{-1}$Mpc. 
A significant part of the difference may be due to
the small fraction of baryons ($\sim6\%$ by $z=3$)
which were in 
the highest peaks and
have condensed out of the IGM to form stars irreversibly
in the simulation.

Figure 2 shows the flux correlation function 
$\xi_{flux} (v)$ [defined by equation (2)] 
for S/N=50, FWHM=6km/s (solid curve) and
S/N=10, FWHM=30km/s (dashed curve). 
We see that the flux correlation depends
on the observational resolution at small separations,
but does not depend on 
the observational resolution at large separations.
For example,
for the two cases with the adopted observational parameters,
the correlation strengths at
$v=(10,100)$km/s 
are 
$\xi_{flux}=(0.20,0.045)$
and 
$(0.16,0.042)$, respectively.

Figure 3 shows the ``capped optical
depth correlation function" 
$\xi_{\tau,c} (v)$ [defined by equation (3)] 
for two sets of observational parameters,
S/N=50, FWHM=6km/s (thick curves) and
S/N=10, FWHM=30km/s (thin curves).
For each case three functions for three
optical depth ceiling values are shown:
$\tau_{max}=\infty$ (solid curve),
$\tau_{max}=4.0$ (dotted curve)
and
$\tau_{max}=2.0$ (dashed curve).
Three features are to be noted.
First, similar to the flux correlation function 
($\xi_{flux}$, shown in Figure 2),
this optical depth correlation function
only weakly depends on S/N and FWHM
on scales $\ge 40$km/s.
Second, on scales $\le 40$km/s its values
start to depend more strongly on 
S/N and FWHM,
presumably due to the dominance of small, dense
regions at small scales,
which become under-resolved with lower observational
resolution and/or sensitivity.
Third, the correlation depends strongly
on the imposed ceiling value;
a lower ceiling value 
results in a weaker correlation.
In other words, high optical depth, high density regions
are more strongly clustered than low optical
depth, low density regions on all scales probed here, $1-320$km/s,
with the trend that the difference is progressively larger
on smaller scales.
Note that one side of our simulation box spans a velocity
range of 1280km/s, so the reader is reminded that
the correlation computed here is only valid
for scales $\le 320$km/s (a quarter of the box size),
because of the periodic boundary condition.
It is noted that
the two-point correlation function measured by
any of the proposed functions here
is underestimated due to the missing
density waves larger than our
simulation box ($\lambda > 10h^{-1}$Mpc).
Since $\xi(r)\propto P_k k^2 sin(kr) d\ln k$ 
and $P_k\sim k^{-2}$ at $\lambda=10h^{-1}$Mpc
(and the fiducial power index becomes progressively larger than
-2.0 on larger scales) for the adopted model,
we expect that the underestimate
is of a logarithmic factor, perhaps a factor less than two
for the computed correlations on scales $v\le 300\;$km/s.

To see separately 
the effects of thermal broadening and peculiar velocity
we show, in Figure 4,
the corresponding correlations 
with only the thermal broadening effect (dotted curves),
with only the peculiar velocity  effect (dashed curves)
and with both effects (solid curves; the same as the thick curves in
Figure 3)
for the case with
S/N=50, FWHM=6km/s.
For each case three functions for three
optical depth ceiling values are shown:
$\tau_{max}=\infty$ (thick curve),
$\tau_{max}=4.0$ (medium curve)
and
$\tau_{max}=2.0$ (thin curve).
We see that the peculiar velocity of the gas
serves to boost correlations on both large ($v\ge 100$km/s)
and small scales ($v<10$km/s),
but to suppress correlations on intermediate scales.
The boost of correlation on small scales by the peculiar velocity
produces velocity caustics and the boost on large scales
due to peculiar velocity is essentially the "Kaiser Effect"
(Kaiser 1987).
Since the total number of pairs has to be conserved,
the correlation strength
on intermediate scales is reduced.
The thermal motion alone 
tends to reduce correlations on all scales.
However, coupling of peculiar velocity
and thermal motion is not simple.
It appears that the 
correlation is slightly stronger at the intermediate
scales $v=15-60$km/s when both effects are included
than that with either of the effects alone.
This result reflects the complicated
interplay between thermal motion and peculiar velocity
of the gas on the correlation.
The result that the correlation function at small
scales is reduced when both the thermal and
velocity effects are included
indicates that 
velocity caustics are largely removed 
by the thermal broadening effect.
But overall, the two effects do not seem to significantly alter
the true correlation of the clouds on scales $\ge 20$km/s.

Figure 5 shows the ``stepped optical depth correlation function"
$\xi_{\tau,s} (v)$ [defined by equation (6)] 
for S/N=50, FWHM=6km/s (thick curves) and
S/N=10, FWHM=30km/s (thin curves).
For each case three functions for three
optical depth floor values are shown:
$\tau_{min}=1.0$ (solid curve),
$\tau_{min}=2.0$ (dotted curve)
and
$\tau_{min}=4.0$ (dashed curve).
We remind the reader that a higher value of 
$\tau_{min}$ means a selection of only higher optical depth regions.
We see that, at $\tau_{min}=1.0$ (solid curves),
observational parameters S/N and FWHM do not 
affect the results on the scale $\ge 30$km/s, 
simply because the dominant contributions from regions of 
optical depths just above the floor value
are resolution insensitive.
Equivalently stated,
these regions are already well resolved even by
the quoted lower resolution observations.
The situation is somewhat different for higher 
$\tau_{min}$'s, where higher S/N and lower FWHM give
slightly weaker correlations.
This is because lower S/N and higher FWHM observations
only pick out relatively higher density, high optical
depth regions,  which are intrinsically more strongly clustered.
On small scales, the observational resolution
dependence is seen for all three 
$\tau_{min}$'s, where higher S/N and lower FWHM give
stronger correlations as expected,
which is, of course, 
directly related to the greater resolving power of the relevant
dense regions in high resolution observations.
Consistent with results shown 
in Figure 3, 
we find here that high optical depth, high density regions
are more strongly clustered than lower optical
depth, low density regions on all scales probed here, $1-320$km/s.

Similar to Figure 4,
Figure 6 shows separately the correlations $\xi_{\tau,s}$
with only thermal broadening (dotted curves),
with only the peculiar velocity  effect (dashed curves)
and with both effects (solid curves; the same as the thick curves in
Figure 5).
For each case three functions for three
optical depth ceiling values are shown:
$\tau_{min}=4.0$ (thick curve),
$\tau_{min}=2.0$ (medium curve)
and
$\tau_{min}=1.0$ (thin curve).
The reader is reminded that the
main difference between $\xi_{\tau,c}$ shown in Figures (3,4)
and $\xi_{\tau,s}$ shown in Figures (5,6)
is that $\xi_{\tau,c}$ measures the correlation by excluding
high density, high optical depth regions,
while $\xi_{\tau,s}$ measures the correlation by excluding
low density, low optical depth regions.
We see that
the correlation with peculiar velocity alone has the
highest amplitude, the one with thermal motion alone
has the lowest 
amplitude, while the correlation with both effects
is intermediate at all scales except $v>100$km/s. 
This result is consistent with the fact that
high density regions are more strongly clustered and
peculiar velocities tend to enhance the correlation strength
by, on average, decreasing the pairwise distance in velocity space
(Kaiser 1987).

Finally, we show in Figure 7
the line-line correlation function  
$\xi_{line} (v)$ [defined by equation (7)] 
for S/N=50, FWHM=6km/s (thick curves) and
S/N=10, FWHM=30km/s (thin curves),
with a flux threshold $F_{t}=0.6$ (see MCOR).
For each case three functions for two
equivalent widths are shown:
$W_t=10$km/s (solid curve),
and $W_t=40$km/s (dashed curve).
We stress that the correlations on scales $\le 100-200$km/s
are probably sensitive to
the details of the line identification scheme.
In particular, we believe that the sharp downturns of the
line correlation functions
on scales $\le 100-200$km/s are, in large part,
likely due to the blending of lines adjacent in velocity space.
However, the correlations on scales $\ge 200$km/s
are perhaps more robust and less sensitive to 
specifics of a line identification method.
We therefore focus on scales $200-320$km/s, where
the upper limit, corresponding
to a quarter of the box size, 
arises from the fact that our simulation box
assumes periodic boundary conditions.
We see that a higher $W_t$
results in a stronger correlation on scales $\ge 200$km/s. 
This trend is consistent with observations that
there is a gradual progression in clustering strength
from weak to strong with increasing column density
(Crotts 1989; Cristiani \etal 1997).
The results are shown to be insensitive
to S/N and FWHM.
To examine the dependence of the line-line correlation
function on the flux threshold $F_t$, 
Figure 8 shows $\xi_{line} (v)$ 
at $F_t=0.4$ (solid curve), $0.6$ (dotted curve)
and $0.8$ (dashed curve),
each with S/N=50, FWHM=6km/s and $W_t=10$km/s.
We see that demanding a lower $F_t$ is 
qualitatively equivalent to demanding a larger
equivalent width $W_t$, as shown in Figure 7.

The above separately
displayed results 
are best presented as summarized in Table 1,
where we show the clustering strength of all the correlation
measures at separations of $50,100,200$ and $300$km/s (at $z=3$).
All measures use S/N=50, FWHM=6km/s.
For $\xi_{line}$, $W_t=10$km/s is used 
but we note that results with $W_t=40$km/s
are very similar.
In addition, two measures for galaxy
clustering are shown:
``$\xi_{galaxy}$" indicates 
the galaxy-galaxy correlation strength at $z=3$
if we assume that the galaxies are (linearly) biased over 
dark matter by an amount equal to that at $z=0$, i.e.,
$b\equiv \sigma_8^{-1}=1.27$, where $\sigma_8$ is the density
fluctuation in a top-hat sphere of radius $8h^{-1}$Mpc at $z=0$;
$\xi_{stable}$ is obtained
by extrapolating the observed $z=0$ galaxy-galaxy two-point 
correlation function [$\xi(0)=(r/5.5h^{-1}\hbox{Mpc})^{-1.8}$]
to $z=3$ assuming stable clustering
(i.e. correlation function is fixed in real space).
A few points become clear from the table.

First, the cosmological model adopted 
is interesting in that ``$\xi_{galaxy}$" 
is in agreement with $\xi_{stable}$ at $z=3$.
However, if galaxies at $z=3$ turn out to be much more strongly
clustered than that indicated by $\xi_{galaxy}$,
then a higher bias of galaxies over matter is
required (if one adopts this particular cosmological model)
and it implies that the stable clustering assumption is invalid,
vice versa.
Note that this model also reproduces all the observations at $z=0$
quantitatively well.
Second, the flux correlation $\xi_{flux}$ does not
represent the true matter correlation well,
partly because of its definition 
and partly because of the fact that it does not particularly
reflect the correlation of dense regions (i.e. ``clouds").
Third, the optical depth correlation directly using
the observed optical depth 
[$\xi_{\tau,c}(\infty)$]
traces the true correlation of the underlying matter $\xi_{\rho_d}$
well with some moderate anti-bias at scales $\le 100$km/s.
Removing high optical depth regions significantly reduces
the correlation on all scales; 
$\xi_{\tau,c}(2)$ is about a factor of two smaller than
$\xi_{\tau,c}(\infty)$ at all scales.
Fourth, the best representation of the matter
correlation is achieved by $\xi_{\tau,s}(2)$,
which faithfully traces $\xi_{\rho_d}$ on all the
scales shown. Since optical depth $\tau=2$ should be
easily identifiable with the current observational
sensitivity without much distortion caused by noise,
$\xi_{\tau,s}(2)$ may represent the best correlational measure
both to compare observations with simulations and 
to compare correlations of $\lya$ clouds with 
correlations of galaxies.
We note that
a $\lya$ cloud with a column density 
of $N=1\times 10^{14}$cm$^{-2}$
and a Doppler width $b=25$km~s$^{-1}$ has a central
$\lya$ optical depth of $\sim 3.0$.
Therefore, 
$\xi_{\tau,s}(2)$ measures the spatial
correlation of $\lya$ clouds 
with column densities $\ge 10^{13-14}$/cm$^2$.
A $\tau_{min}$ less than $2.0$ enables 
us to compute correlations of lower column density clouds.
The essential difference between the optical
depth based correlation measures and flux based measures
is that optical depth directly probes the densities of the clouds,
thus a direct measure of density.
Finally, we emphasize that the line-line correlation 
is highly dependent on the details of the line
identification scheme.
While a comparison between observations and simulations can always 
be made as long as both are analyzed in the same way using
a chosen measure,
the difference between the line-line correlational measure,
which is almost the sole, used clustering
indicator of observed Lyman forest correlations,
and $\xi_{\tau,s}(2)$, which represents an alternative to the 
line-line correlation
measure, seems clear.
The number of lines identified for a given ``true" cloud 
(which is easy to see in simulations) 
or a given domain of observed spectra
depends on observational resolution
and noise as well as post-observation
fitting procedures. Even in the limit of infinite observational sensitivity
the number of lines determined is not unique,
due to the ambiguity in the line fitting and deblending process.
We note that
$\xi_{\tau,s}$ picks out high optical depth
regions,
gives each (uniform) spectral pixel in the selected
region equal weight, and measures the correlation of these pixels.
In essence, $\xi_{\tau,s}$ 
is a line-line correlation, if one visualizes each selected spectral
pixel as a ``line",
but it lacks any ambiguity.

While a direct comparison with available observations is
not attempted due to the lack of a common measure,
we see a broad consistency.
Sargent \etal (1980)
find that there is very weak correlation on scales
$300-30,000$km/s with an observational FWHM$\sim 50$km/s 
at $z=1.7-3.3$.
Webb (1986) finds a correlation strength of $0.32\pm 0.08$
for $\Delta v=50-290$km/s with FWHM$\sim 20$km/s at $z=1.9-2.8$.
Meiksin \& Bouchet (1995) find 
positive correlations, with a maximum amplitude of $0.5-1.0$,
on the scales $\Delta v=100-600$km/s with FWHM$\sim 25$km/s
at $z=2-4$.
Rauch \etal (1992) show that 
there is no strong clustering on all scales probed,
$\Delta v > 100$km/s with FWHM$\sim 20$km/s at $z=2.7-3.4$.
Cristiani \etal (1997) conclude that
there is significant clustering at 
$\Delta v < 300$km/s with FWHM$\sim 10$km/s at $z\sim 3.0$,
with $\xi \sim 0.6$ at $\Delta v = 100$km/s for 
clouds with $N_{HI}\ge 10^{13.8}$cm$^{-2}$,
and no significant clustering for 
$N_{HI}\le 10^{13.6}$cm$^{-2}$.

\section{Discussion and Conclusions}

Examination of several measures of the correlation of $\lya$ clouds
and comparison between them and fundamental matter correlations
shows that mass is more strongly clustered
than that of $\lya$ clouds, on all scales at $z=3$.
It is also found that, among the $\lya$ clouds,
higher density, higher optical depth, higher column density
regions are more strongly
clustered than lower density, lower optical depth, lower column density
regions,
with the difference being larger at small separations and
smaller at large separations.
These effects should be observable even in the current observations.
Although we are not able to make
a direct comparison with available observations 
due to the lack of a common measure,
we see a broad consistency between
our simulation results and observations.
If we assume that galaxies form preferentially
in regions whose densities are much higher than
those of the $\lya$ clouds,
especially at the redshift considered here,
then a consistent picture emerges:
the correlation strength for a given set of objects is positively 
correlated with their characteristic global density
and the differences among the correlations of galaxies, $\lya$ clouds
and mass reflect the differences in density that each trace.
The property might be tracable to the nature
of Gaussian density field where higher density peaks are
more strongly clustered than lower density peaks (Bardeen \etal 1986).

The true correlation strength of the matter 
is about $0.2-2.0$ on scales $50-300$km/s,
where it can be measured reliably in our simulations, at $z=3$.
Significant positive correlations with a 
strength of $0.1-1.0$ are predicted for $\lya$ clouds
in the velocity range $50-300$km/s.
All correlation functions examined here
decrease monotonically with increasing separation.
Due to our limited box size ($10h^{-1}$Mpc)
the computed correlation
underestimates the true correlation.
However, the underestimate 
is expected to be a factor less than two at $v \sim 300\;$km/s
and smaller at smaller scales.
While our limited simulation box size hampers our
ability to study the correlations of the $\lya$
forest on scales $\ge 2.5h^{-1}$Mpc ($\Delta v=320$km/s),
we expect that the correlation strength of $\lya$ clouds
on scales $\ge 300$km/s is small, $\le 0.1$.
The correlation strength on scales of several hundred km/sec
measures the power of the density fluctuations 
on comoving megaparsec scales.
Thus, an accurate determination of
the correlation strength on scales of several hundred km/sec
provides a unique data point for the power spectrum 
on comoving megaparsec scales.

Among the measures examined, the ``stepped optical depth correlation function"
(Equation 5) seems to be the most useful correlational measure.
It reasonably represents the true correlation of the underlying
matter, thus enabling a better indication
of both matter correlation and the relationship between galaxies
and $\lya$ clouds.
Moreover, it appears to provide a simple
alternative to the conventional line-line correlation function
with a few obviously advantageous features: 
1) it does not require ambiguous post-observation
line fitting and deblending 
procedures such as those commonly employed 
in the conventional line finding methods,
2) it does not depend sensitively on the observational resolution
such as FWHM, once a ``true" cloud (which is hard to define
observationally but easy to identify and study 
in three-dimensional simulations) is resolved (this typically means
that an FWHM is sufficiently smaller than the thermal broadening
width of a cloud) and,
3) it can be easily measured with the current observational
sensitivity without being contaminated significantly by the
presence of noise, if one chooses an appropriate optical
depth floor value $\tau_{min}$,
say, $\le 2.0$.

Applying the newly proposed correlational measures, particularly, 
$\xi_{\tau,s} (v)$ [defined by equation (5)],
to observed spectra, 
both along the line of sight
and in the perpendicular direction for nearby sightlines,
will perhaps yield much more unambiguous 
results about the spatial clustering of the observed $\lya$ clouds.
Comparison of correlations between observed
and model spectra employing such a simple unambiguous
measure might enable a much simpler interpretation
of some essential issues such as if the clouds
are made of smaller dense cloudlets.

Aside from the choice of a particular cosmological model which
seems viable,
the critical ingredient in our simulation is
the assumption that 
the ionizing radiation field is spatially uniform.
While it seems reasonable,
future larger simulations with box size 
$\ge 20h^{-1}$Mpc are required in order to study 
correlations up to the scale $5h^{-1}$Mpc.
Such a study will shed light on the origin
of the correlations of $\lya$ forest and possibly
also on the properties of the meta-galactic UV background,
when compared with observations.

\acknowledgments
The work is supported in part
by grants NAG5-2759, AST91-08103 and ASC93-18185.

\newpage
\figcaption[FLENAME]{
shows 
$\xi_{\rho_b} (r)$ (solid curve)
and
$\xi_{\rho_d} (r)$ (dashed curve).
Note that the two correlations 
are computed in real space as a function of proper distance,
as shown in the bottom horizontal axis.
For reference velocity distance, converted from
the proper distance using the relation $v=r/H(z)$,
where $H(z)=512$km/s/Mpc is the Hubble constant at $z=3$
for the adopted model, is shown in the top horizontal axis.
Note that $1+\xi$ rather than $\xi$ is shown in the plot
in order to display negative values of $\xi$ in the logarithmic scale.
\label{fig1}}

\figcaption[FLENAME]{
shows the flux correlation function 
$\xi_{flux} (v)$ [defined by equation (2)] 
for two sets of observational parameters,
S/N=50, FWHM=6km/s (solid curve) and
S/N=10, FWHM=30km/s (dashed curve). 
\label{fig2}}

\figcaption[FLENAME]{
shows the ``capped optical
depth correlation function"
$\xi_{\tau,c} (v)$ [defined by equation (3)] 
for two sets of observational parameters 
S/N=50, FWHM=6km/s (thick curves) and
S/N=10, FWHM=30km/s (thin curves).
For each case three functions for three
optical depth ceiling values are shown:
$\tau_{max}=\infty$ (solid curve),
$\tau_{max}=4.0$ (dotted curve)
and
$\tau_{max}=2.0$ (dashed curve).
\label{fig3}}

\figcaption[FLENAME]{
shows the ``capped optical depth correlation function"
$\xi_{\tau,c} (v)$ [defined by equation (3)] 
with only thermal broadening (dotted curves),
only the peculiar velocity effect (dashed curves)
and both effects (solid curves; the same as the thick curves in
Figure 3)
for the case with
S/N=50, FWHM=6km/s.
For each case three functions for three
optical depth ceiling values are shown:
$\tau_{max}=\infty$ (thick curve),
$\tau_{max}=4.0$ (medium curve)
and
$\tau_{max}=2.0$ (thin curve).
\label{fig4}}

\figcaption[FLENAME]{
shows the ``stepped optical depth correlation function"
$\xi_{\tau,s} (v)$ [defined by equation (5)] 
for two sets of observational parameters 
S/N=50, FWHM=6km/s (thick curves) and
S/N=10, FWHM=30km/s (thin curves).
For each case three functions for three
optical depth floor values are shown:
$\tau_{min}=1.0$ (solid curve),
$\tau_{min}=2.0$ (dotted curve)
and
$\tau_{min}=4.0$ (dashed curve).
\label{fig5}}

\figcaption[FLENAME]{
shows the ``stepped optical depth correlation function"
$\xi_{\tau,s} (v)$ [defined by equation (5)] 
with only thermal broadening (dotted curves),
only the peculiar velocity  effect (dashed curves)
and both effects (solid curves; the same as the thick curves in
Figure 5).
For each case three functions for three
optical depth ceiling values are shown:
$\tau_{min}=4.0$ (thick curve),
$\tau_{min}=2.0$ (medium curve)
and
$\tau_{min}=1.0$ (thin curve).
\label{fig6}}

\figcaption[FLENAME]{
shows the line-line correlation function  
$\xi_{line} (v)$ [defined by equation (7)] 
for two sets of observational parameters 
S/N=50, FWHM=6km/s (thick curves) and
S/N=10, FWHM=30km/s (thin curves),
with a flux threshold $F_{t}=0.6$ (see MCOR).
For each case three functions for two
equivalent widths are shown:
$W_t=10$km/s (solid curve),
and $W_t=40$km/s (dashed curve).
\label{fig7}}

\figcaption[FLENAME]{
shows $\xi_{line} (v)$ 
at $F_t=0.4$ (solid curve), $0.6$ (dotted curve)
and $0.8$ (dashed curve),
each with S/N=50, FWHM=6km/s and $W_t=10$km/s.
\label{fig8}}

\clearpage
\begin{deluxetable}{ccccccccccc} 
\tablewidth{0pt}
\tablenum{1}
\tablecolumns{10}
\tablecaption{Summary of correlations} 
\tablehead{
\colhead{\vbox{\hbox{\ \ $\Delta v$}\hbox{(km/s)}}} &
\colhead{\vbox{\hbox{\ \ \ $\Delta x$}\hbox{($h^{-1}$Mpc$^a$)}}} &
\colhead{\bf $\xi_{\rho_d}$} &
\colhead{$\xi_{\rho_b}$} &
\colhead{$\xi_{flux}$} &
\colhead{$\xi_{\tau,c}(\infty)$} &
\colhead{$\xi_{\tau,c}(2)$} &
\colhead{\bf $\xi_{\tau,s}(2)$} &
\colhead{$\xi_{line}$} &
\colhead{``$\xi_{galaxy}$"$^{c}$} &
\colhead{$\xi_{stable}$$^{d}$} }

\startdata
$50$ & $0.39$ & {\bf $1.6$} & $1.1$ & $0.10$ & $0.95$ & $0.43$ & {\bf $1.58$} & $-0.35^{b}$ & $2.0$ & $1.83$ \nl
$100$ & $0.78$ & {\bf $0.70$} & $0.40$ & $0.050$ & $0.34$ & $0.17$ & {\bf $0.57$} & $0.15^{b}$ & $0.89$ & $0.52$ \nl
$200$ & $1.56$ & {\bf $0.20$} & $0.20$ & $0.007$ & $0.16$ & $0.06$ & {\bf $0.16$} & $0.36$ & $0.25$ & $0.15$ \nl
$300$ & $2.34$ & {\bf $0.20$} & $0.20$ & $0.005$ & $0.06$ & $0.03$ & {\bf $0.10$} & $0.11$ & $0.25$ & $0.073$ \nl
\enddata
\tablenotetext{a}{comoving length units}
\tablenotetext{b}{probably underestimated due to line blending}
\tablenotetext{c}{galaxy correlation at $z=3$ derived from $\xi_{\rho_d}$ assuming a linear density bias of $1.27=1/\sigma_8$}
\tablenotetext{d}{galaxy correlation at $z=3$ derived by extrapolating present observed galaxy correlation function to $z=3$ assuming stable clustering}
\end{deluxetable}

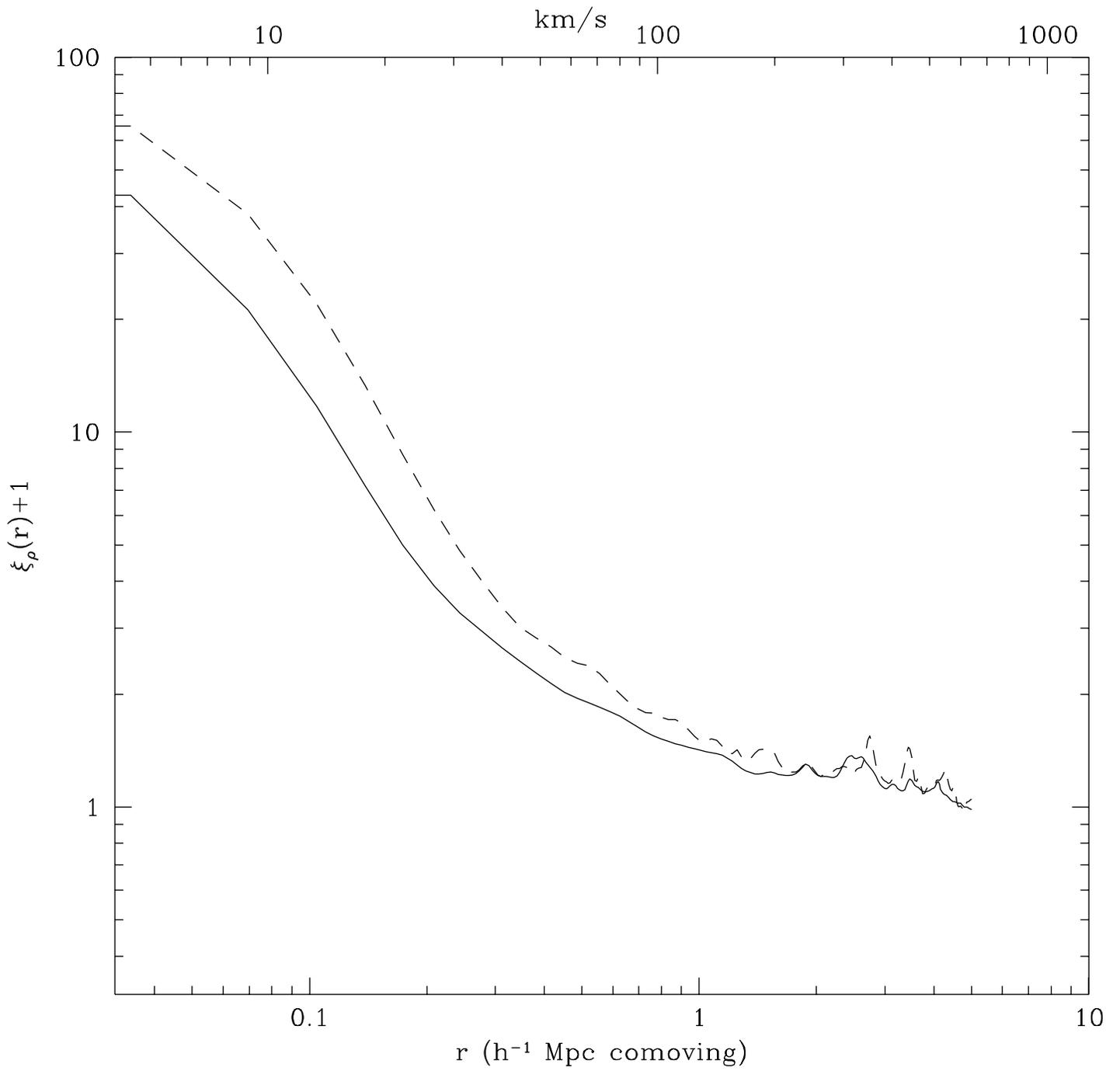
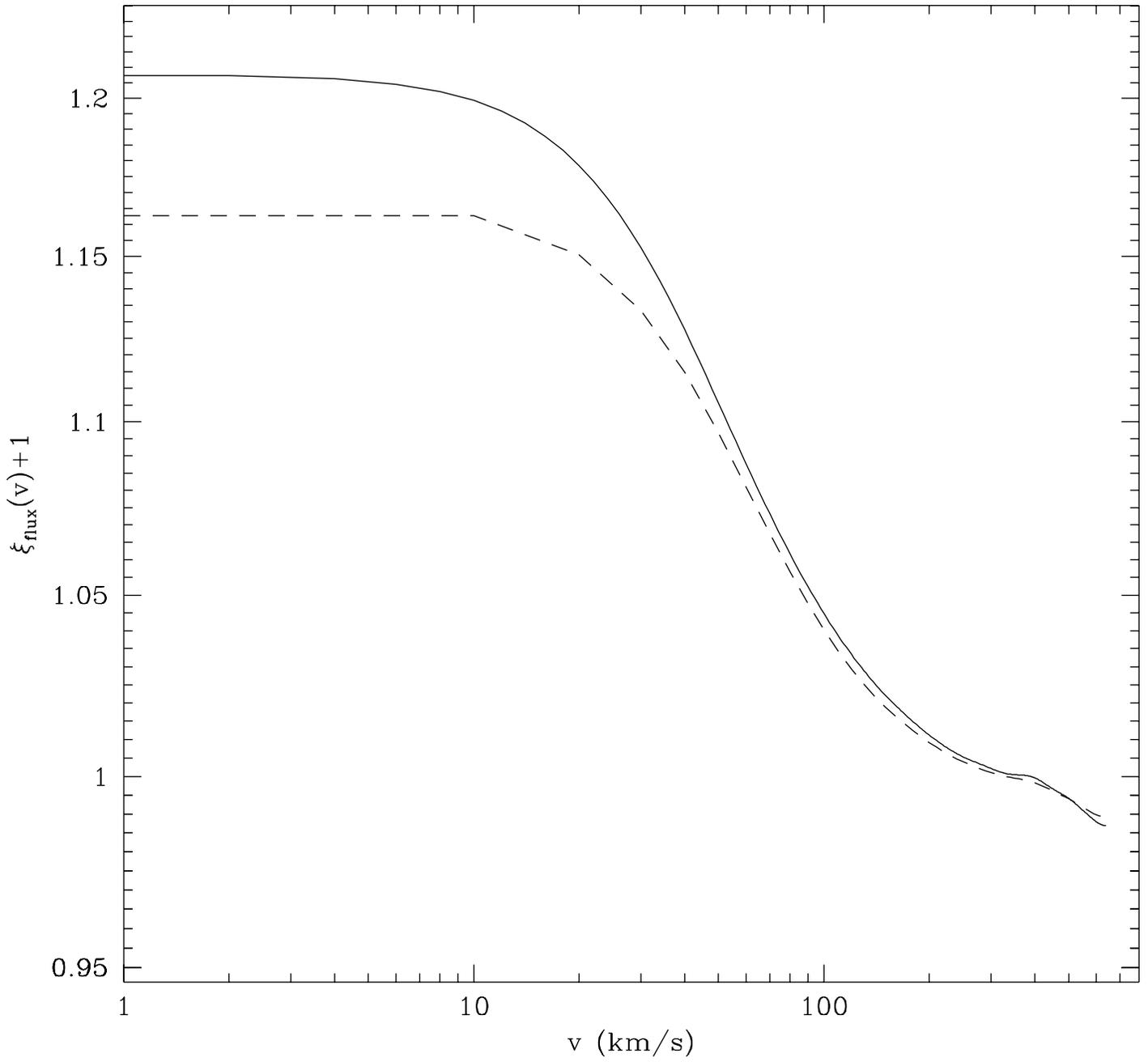
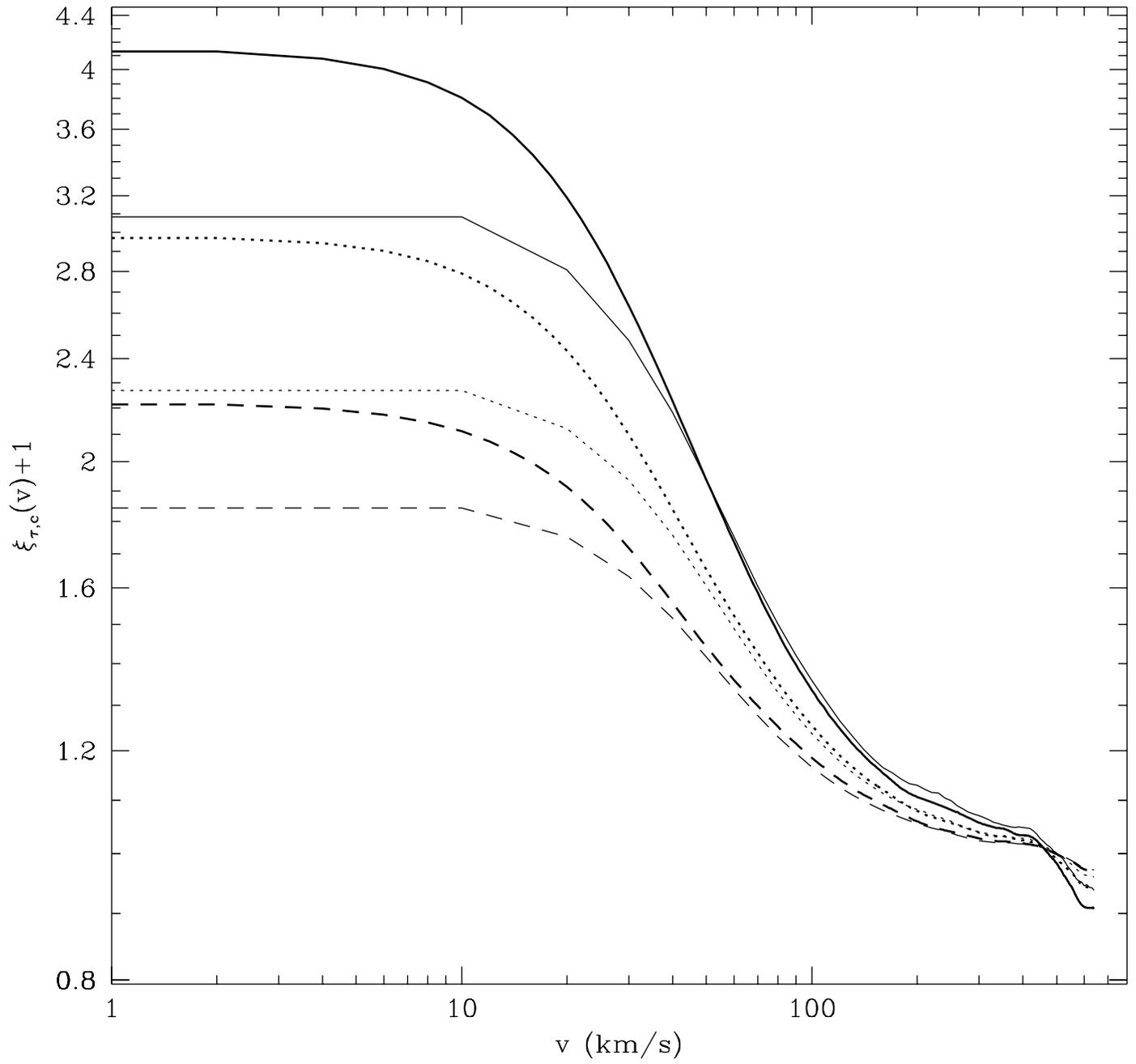
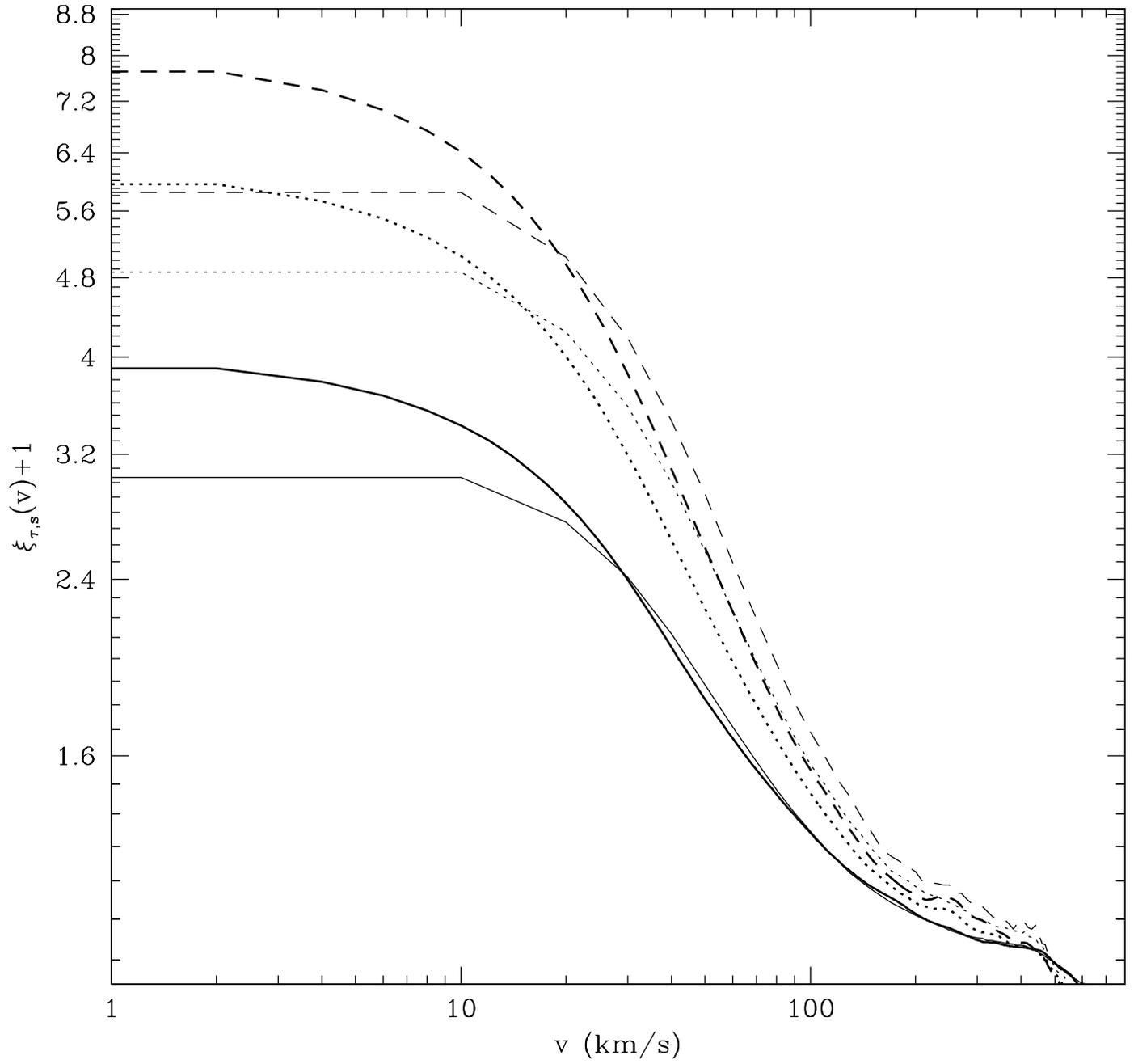
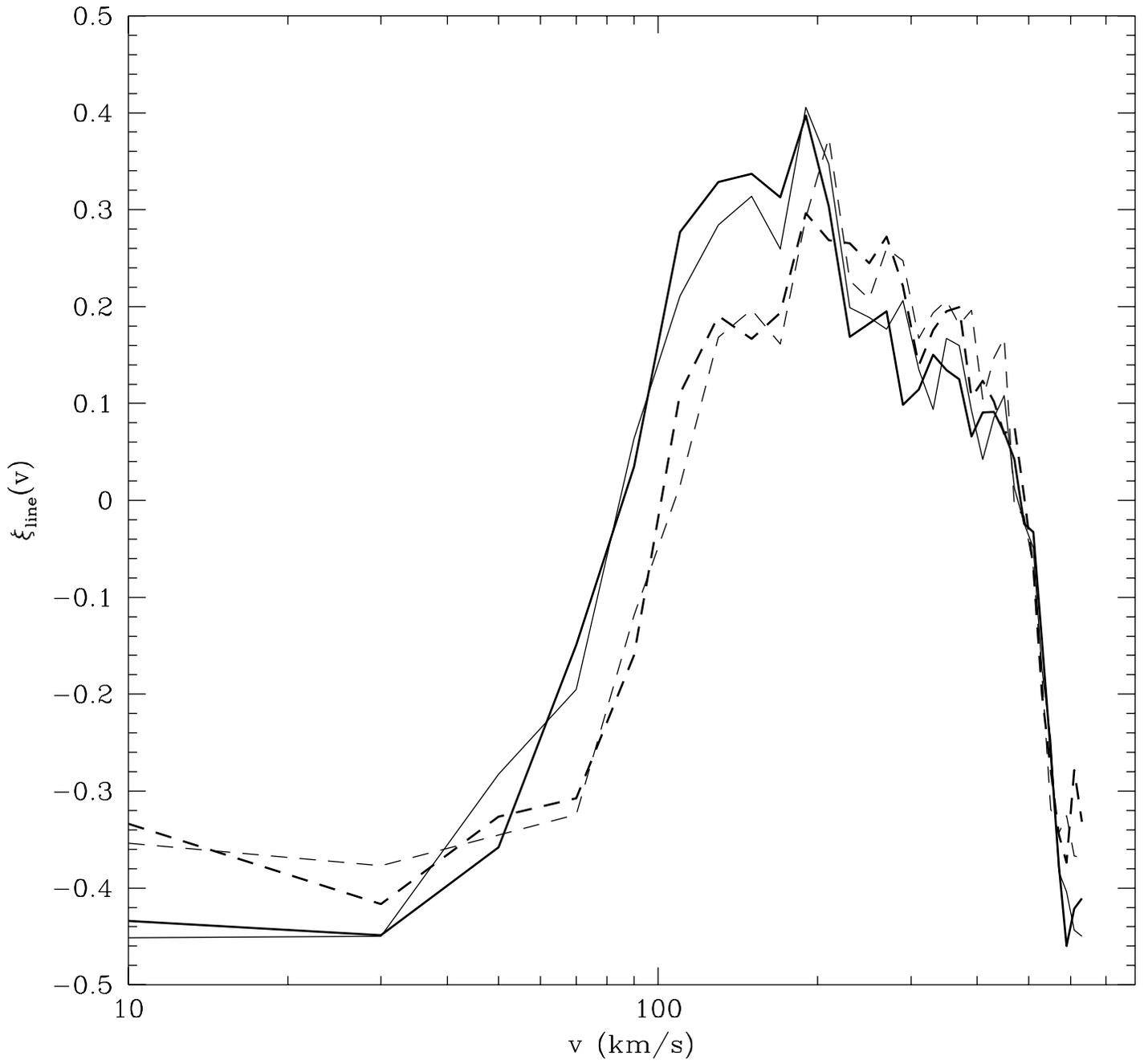
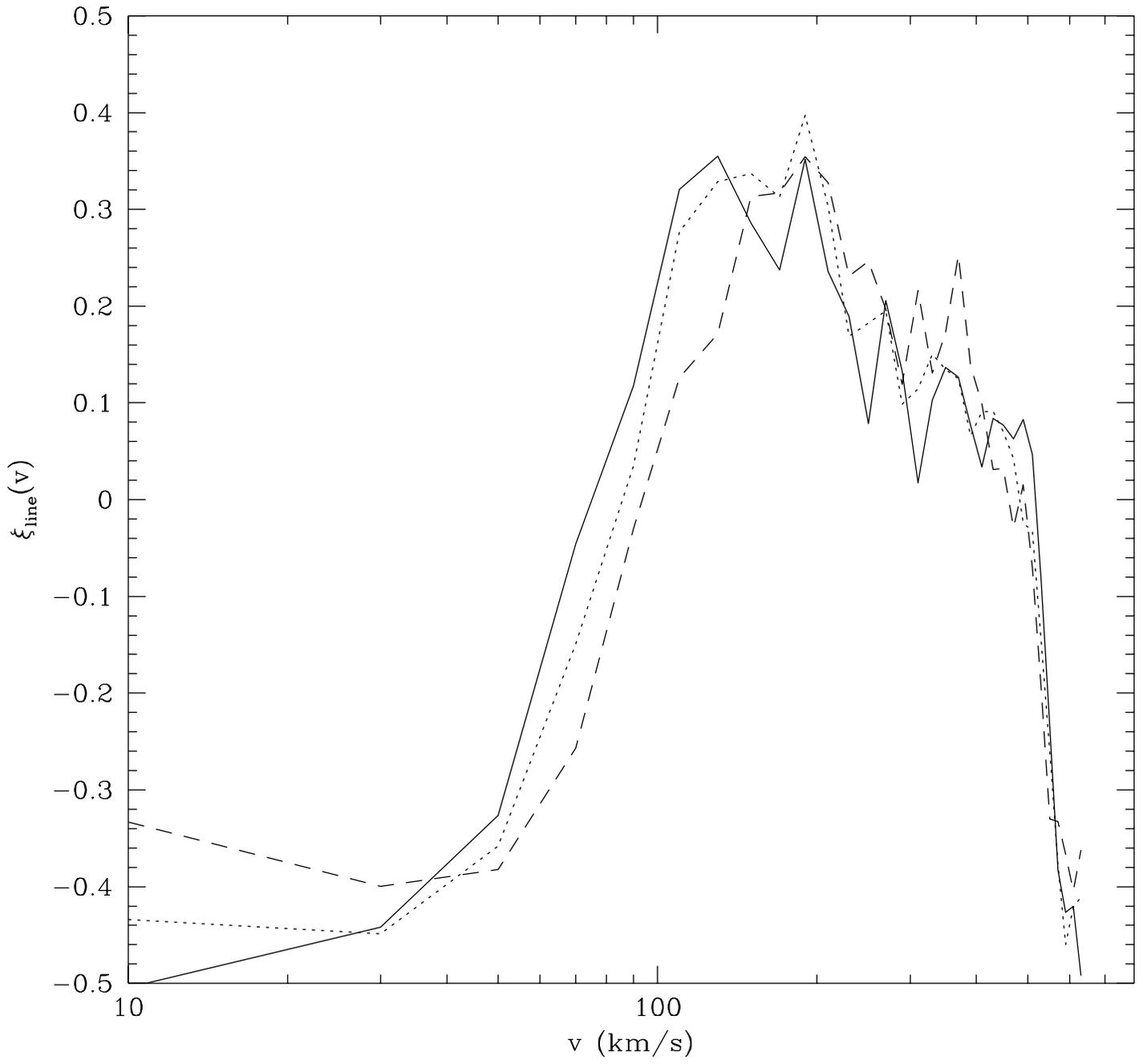
\end{document}